\title{A static Theory of Promises}
\author{Jan A. Bergstra\footnote{%
Email: \texttt{j.a.bergstra@uva.nl,janaldertb@gmail.com}.}
\\Informatics Institute, University of Amsterdam\\
Science Park 904, 1098 XH Amsterdam\\
The Netherlands\\~\\
Mark Burgess\footnote{%
Mark Burgess was supported by the 
EC IST-EMANICS Network of Excellence (\#26854). 
Email: \texttt{mark.burgess@cfengine.com}}\\
Visiting Informatics Institute, University of Amsterdam}

\documentclass[12pt]{article}
\date{}

\usepackage{alltt}
\usepackage{epsfig}
\usepackage{a4}
\usepackage{times}

\def\beq{\begin{eqnarray}}
\def\eeq{\end{eqnarray}}
\def\2{\frac{1}{2}}

\newtheorem{definition}{Term}

\begin{document}
\newcommand{\promise}[1]{\stackrel{#1}{\longrightarrow}}

\maketitle

\begin{abstract}
We discuss the concept of promises within a framework that
can be applied to either humans or technology.  We compare promises to
the more established notion of obligations and find promises to be
both simpler and more effective at reducing uncertainty about multi-agent behaviour.
\end{abstract}
\tableofcontents
\section{Introduction}

This paper\footnote{%
This paper is a revision of the paper arXived on 
\texttt{http://arxiv.org/abs/0810.3294v4}.
In this version remarks on the institution of promising in the context of 
Philosophy of Law have been added, as well as a 
reference to work on the anthropology of promising. Various small mistakes were remedied as well.}
is about the usefulness and importance of promises as a
concept.  For something that is so abundant in daily life, the concept
of a promise has been given only a limited treatment in
academic literature, and then mainly in the areas of philosophy, law, and
economics. Promises give way more often to the idea of obligations,
indeed many authors quickly do away with promises and replace them
with apparent obligations as if this were a necessity. No doubt this
tendency originates in historical matters, but it is unfortunate as
promises have a plausible and practical value both in their mundane
meaning for humans and in an extended interpretation that can apply to
machines and other inanimate objects.

The concept of a promise is not a difficult one -- in fact we claim it to be a
simpler concept than an obligation -- but it requires some
care, and it leads to some complications. 
In this paper we present our understanding of promises and
explain why in some application areas 
there are both practical and theoretical advantages to
their use over obligations. To make our case unambiguously, we ask the
forbearance of readers in presenting the fundamentals and motivations
at some length.

We assume the existence of an emerging topic of promise theory, to which
we hope to make a contribution with our paper. The development of promise theory
is stil in an initial phase.
We position promise theory as a branch of informal logic. We 
mention~\cite{Johnson2006,Hamby2012,Blair2013} 
and~\cite{Groarke2013}
as introductions to informal logic and~\cite{YuanK2011} for an example of 
how informal logic may be applied in computing. Promise theory has been specifically developed with
applications in distributed systems in mind, but not limited to applications in informatics.

Promise theory may therefore be viewed alternatively as a topic in theoretical informatics 
with foundations in informal logic and with a potential for application in management science.
Promise theory presupposes a theory of agents. It draws upon the classical notion of reciprocity,
that is agents modeling one another recursively (see~\cite{GmyD1993}). For the notion of an
agent we refer to~\cite{PanzarasaJ2001}.

\section{A motivation}

For surely all readers the notion of a promise will be quite familiar.
Our experience of promises might be both positive and negative, for
many promises that are made are never kept. Below are examples of the
kinds of statements we intend to call promises. Let us begin with everyday
statements and progress gradually to the kinds of abstract promises
that we would like to use in a variety of technical scenarios.

\begin{itemize}
\item I promise you that I will walk the dog.
\item I promise you that I fed your cat while you were away.
\item We promise to accept cash payments.
\item We promise to accept validated credit cards.
\item I promise to lock the door when I leave.
\item I promise not to lock the door when I leave.
\end{itemize}

These examples are quite uncontroversial. They are easily found in
every day life, spoken by humans or posted on signs.  We now want to
argue that it is useful to extend the notion of promises to inanimate
objects that have been designed or programmed to behave in a certain
manner. This is not a very large step, but it is easier to make with
some examples.

Consider the following promises that might be made in the world
of Information Technology:

\begin{itemize}
\item The Internet Service Provider promises to deliver broadband Internet at a fixed speed 
for a fixed monthly payment.
\item The security officer promises that the system will conform to security requirements.
\item The support personnel promise to be available by pager 24 hours a day.
\item Support staff promise to reply to queries within 24 hours.
\end{itemize}
Again these are straightforward promises, which could be described
further to be more specific.  The final promise could also be
restated in more abstract terms, transferring the promise to an
abstract entity: ``the help desk'':

\begin{itemize}
\item The company help-desk promises to reply to service requests within 24 hours.
\end{itemize}

\subsection{Promises about a piece of technology}
This latter example illustrates the way that we transfer the
intentions of promises to `entities' that we consider to be
responsible by association. It is a small step from this transference
to a more general assignment of promises to individual components in a
piece of technology.  For example, we can document the properties of
the following tools and technologies in the spirit of this argument:

\begin{itemize}
\item I am a logic gate and promise to transform a {\tt TRUE} 
signal into a {\tt FALSE} signal and vice versa.
\item I am a command line interpreter and promise to accept input and 
execute commands from the user.
\item I am a router and promise to accept packets from a list of authorized IP addresses.
\item I am a compliance monitor and promise to verify and automatically 
repair the state of the system based on this description of system configuration and policy.
\item I am a high availability server and I promise you service delivery with 99.9999\%
availability.
\end{itemize}

From these example we see that the range of application of promises is quite
general. Indeed such promises are all around us in everyday life, both
in mundane clothing as well as in technical disciplines. Statements
about engineering specifications can also profitably be considered as
promises, even though we might not ordinarily think of them in this
way.

When an electronics engineer looks in a component catalogue and sees
`resistors' for sale promising to have resistance of 500 Ohms to
within a tolerance of 5\%, we do not argue about who made this promise
or whether the resistor is capable of independent thought. The
coloured bands on the component are a sufficient expression of this
promise, and we accept it by association. By this reasoning, we
propose that the concept of a promise should be understood in a way
which allows for all of these uses.

\subsection{Range of promiser types}
With the expansion of the use of promise as just advocated goes a rather large range
of types of promisers that we will admit, we mention:
\begin{itemize}
\item Objects (tools, instruments) without intelligence,
\item Non-human animate agents,
\item Intelligent artificial agents,
\item Human agents.
\end{itemize}
The definitions below are intended to apply with intelligent artificial agents\footnote{%
About appropriate definitions of the concept of an (intelligent) agents a significant literature has
been accumulated in the field of artificial intelligence.  } 
acting as promisers
and also in the case of promising human agents. The extension of these definitions to the cases
of objects without intelligence and to non-human animate agents is not carried out in detail,
but the idea is that in such cases an intelligent agent (either human or artificial) is imagined 
which promises on behalf of these objects or agents. The details of the concept of
promising on behalf of another entity will not be considered below.\footnote{%
Developing a definition of promise that applies uniformly for a wide range of promiser types
is left as a conceptual problem for future work.}

\subsection{Promises and reduction of uncertainty}
\label{Reduction-of-uncertainty}
Promises are entities that we use
to form {\em expectations} of the behaviour of all manner of
things. Such expectations contribute to reducing our {\em uncertainty}
about their behaviour, and this can apply as much to technology as to
humans. We therefore take it as given that the concept of a promise is
a useful one, although still in need of a stable definition that is suitable as a basis for promise theory, 
and we consider next how one can formalize promises in the
simplest and least assuming way.

Producing a promise may be more effective in reducing uncertainty than 
putting forward an assertion that is stated with more certainty. 
Indeed if an expectation about a piece
of technology or about an human or artificial agent is asserted with absolute certainty, or merely
with some quantified probability of being valid, the question immediately arises how
that knowledge has been obtained, thereby possibly 
increasing uncertainty rather than reducing it.

Such existential questions do not arise within a community of users for
a piece of technology that has been delivered in combination with a bundle of promises to its user, 
who may simply react to disappointing performance with
a reduction of trust in the promiser. Future promises from that same source will be
received with less optimism. Conversely, if the piece of equipment outperforms the
promised performance that fact may lead to increased trust in the original promiser, 
causing increased confidence in forthcoming promises by the same agent.

\subsection{The logical status of promises: trust/reputation backed action}
Defining what a promise is can hardly be done without examining definitions or
at least approximate descriptions of several similar notions. 
Related but different notions are these:
\begin{itemize}
\item Taking a decision (role backed action).
\item Belief based prediction (a belief backed action).
\item Science based prediction (a science backed action).
\item Announcing a legal claim (legal theory and case history backed action).
\item Announcing a mathematical conjecture (reputation/trust backed action).
\item Specification of properties of an artifact (reputation/trust backed action).
\item Expression of a mathematical fact (a logic backed action).
\item Expression of a quantified expectation (reputation/trust backed action).
\end{itemize}

A promise is an action with a primary and a secondary effect. The primary effect concerns
promisees and other observers who may adapt their actions and plans in accordance 
with becoming aware of the content of a promise. A secondary effect reflects upon a promiser
and impacts on the trust or reputation that the promiser enjoys among those who noticed a promise
being issued. Promising is an action backed by credibility, and trust, the latter mediated by reputation. 

Other forms of backing for an action or an utterance 
exist as well: if a mathematical theorem with its proof is 
communicated and the theorem
is found to lead to inconsistencies with existing theory that fact will 
impact the trust one has in the proof rather than in 
the agent who puts forward the proof. Similarly if a 
scientific claim is found to contradict new findings the 
science that justifies the claim will be scrutinized once more,
rather than its originator.\footnote{Of course repeated findings of problems with mathematical or 
scientific results of the same agent will invoke questions about that agent's credibility,
and so will the discovery of methodological flaws in that agents working processes.
The point we make is that in logic or science backed action the logical or scientific backing of
an action itself can be scrutinized, an option that is absent with trust/reputation backed actions.}

The separation of a notion of promise from related notions like 
prediction (with and without quantified expectation), conjecture, legal claim, 
and decision will require systematic work that depends on the definitions of those other notions. 
That dependency makes it infeasible to analyze these matters in detail within our paper. 
An exception is the notion of a prediction. Predictions can be understood liberally 
(e.g. predictions made by artificial agents) just as promises. 
A prediction differs from a promise in that
its assessment involves no past, present, or future activity or participation of its originator. 
A prediction cannot create an obligation for its creator to comply with the body of the prediction, 
though issuing a prediction may create a responsibility to remedy  expected adverse consequences
of the predicted event or state of affairs. 
Predictions that don't materialize are likely to
be held against the reputation of their originators, however.

\subsection{Extending promise to prediction}
When designing a theory of promises meant for use in informatics some discrepancy with the concept
of promise as used in that theory and the colloquial (or rather philosophical) use of the term 
promise may be acceptable. Once a theory of promises takes a formalized form dedicated to 
a specific application area, such as for instance distributed computing, the emergence of 
a gap between a mainstream informal understanding of promise and its meaning in that 
particular theory is almost unavoidable. With this in mind it may be acceptable in principle that
a theory of promises includes includes predictions under promises as well in spite of the
fact that a meaningful distinction can be made. 

In this way a tour operator can promise (rather than predict) 
a client that a trip next week will profit from good whether conditions, 
\section{The concept of a promise}

Because of their overriding ubiquity, and practical importance, one
would like to have an account of promises that captures their key
properties and explains related concepts such as commitment,
obligation and intention.
There is a surprising lack of discussion about the meaning of promises
in the literature as far as we can tell. Although the concept or its
relatives have been mentioned in such diverse areas from logic, law
and philosophy to economics, information science and computing, there
is no agreement on what constitutes the semantic content of the terms
or if there is even more than a tacit relationship between promise,
commitment, obligation etc. The most attention has been given to the
concept of {\em obligations} especially in the area of deontic logic.
We believe on the other hand that the philosophical implications of
promises are far wider than is generally assumed and that there is
both a need and a practical importance to clarify them once and for
all. Indeed, we will show that  notwithstanding our lengthy definitions 
the concept of a promise is simpler
than that of an obligation.

\subsection{Implicit promise versus explicit promise}
By far the most ubiquitous usage of the term promise is in phrases
like ``the promise of solar energy'' or ``the promise of nano-technology''.
Such promises are implicit in the sense that the existence of a promiser 
is not assumed and the word promise means approximately: expectation
of usability and impact, where an author using the term usually intends
to contribute to an assessment of the mentioned expectation.

In contrast an explicit promise comes about from an action performed
by a promiser in an appropriate context. We will only focus on
explicit promises below and we will assume by default that promises
are explicit.

\subsection{Existing views on promises}
We will survey some views that other authors have put forward on promises,
henceforth understood as explicit promises.

Atiyah~\cite{Atiyah1981} suggests that any promise leads to an obligation
to keep that promise that is motivated by the threat of tit for tat
reprisals.  Reciprocation is thus coupled to the idea of promises
immediately, which seems to hop over fundamental definitions directly
to a discussion of the economics of keeping promises. The obligations
are to avoid injury and to reciprocate goodwill. It might be discussed
whether incentives are the same as obligations. Atiyah points out
however that promising something cannot be necessarily used to create
obligation at will.  Promises might cause obligations but they can
also represent obligations that already exist, i.e. to show commitment
to an existing obligation to pay the price of something.  e.g. I
promise to pay the bearer the sum of 1 pound (in gold). This is only
an existing admission of moral obligation.  Atiyah maintains,
plausibly, that the motivation for promising has changed throughout
history.  When people make promises, their intentions are culturally
bound. Thus a Victorian gentleman's conception of a promise might not
fit with that of a present-day child who promises to be home in time for
dinner.

Cartwright takes Atiyah's view and asks what might be the point of
promises if not to generate the assumed obligation~\cite{cartwright1}.
Why do people bother to make promises about things to which they are
already obliged? His answer includes the idea that it is a face-saving
measure: to mitigate their humility, suggesting that an obligation is
interpreted as a kind of attack or levy of force? Alternatively, perhaps the
obligation to keep one's promises weighs heavier than the original
obligation (I promise you my word as a gentleman not to kill you, even
though the law says I am forbidden).
Referring to Fried~\cite{fried1}, Cartwright points out that the
economics of contractual tit-for-tat suggested by Atiyah is tied
to promises and not to the obligations they might confer.

The idea that promises are an economic driver of contracts or
agreements as bilateral exchanges of promises is continued in the work
of Gilbert~\cite{gilbert1}.  Then Carrillo and Dewatripont have argued
that promises can best be understood as a market mechanism for
reducing the uncertainty in a moral-hazard game~\cite{carrilo1}.  This
work does not seem to have been pursued. Does a promise increase the
likelihood of voluntary cooperation?
A number of other works mention the concept of promises in the context
of game theory also. In these, the concept of a promise is tacitly assumed to
be related to the probability of choosing a particular game strategy.

Scanlon~\cite{scanlon1} meticulously analyses how and to what extent promises give rise
to obligations under a variety of combinations of additional assumptions. In his
analysis morality plays a important role and it is implicitly assumed that 
promiser and promisee are human beings capable of moral reflection.

Zhao et al.~\cite{zhao1} provide a comprehensive modal logic incorporating 
beliefs, capabilities and promises. Unfortunately it is difficult from that work to
extant a clear intuition of the concept of a promise that the authors had in mind. It seems
that this difficulty is in part caused by the formalist approach taken. In
Framinan and Leisten~\cite{framinan1} order promising is displayed as a standard technical term in 
industrial workflow management, while at the same time that use of the term promise
is considered lacking a sufficiently clear definition.

Promises are often treated in the context of agreements, see for instance 
Scheinman~\cite{Scheinman2011}.
Agreements may be constructed out of mutual promises. But promises may exist without
being part of agreements.

\subsection{Promises in distributed computing}
More recently, a different motivation for promises was introduced by
Burgess in the context of distributed
management~\cite{burgessdsom2005}. Rather than focusing on morals or
even economics as the principal motivator, Burgess uses the promise as
a measure of `voluntary cooperation' as a way of circumventing what
we consider to be  fundamental
problems with logics of obligation for determining system behaviour.
Voluntary cooperation is seen as a way of simplifying constraints and avoiding
many-worlds paradoxes.
He pursues the argument further by emphasizing the role of autonomy of
the parts, and argues for a `promise theory' in which every component
in a system that can have unique information or independent action
should be viewed as axiomatically autonomous~\cite{rosegarden}. Any
cooperation or even subordination of the parts that comes about in an
organized system must then be understood as the result of `honouring' purely
voluntary promises to do so. Burgess argues that no matter what one
believes about the power to oblige (even soldiers can refuse to follow
orders), voluntary cooperation can be used as a pragmatic
engineering methodology for mapping out the complexity of a control
problem in a way that is invariant with respect to centralization or
decentralization of systems.

In computer science, particularly the field of Multi-Agent Systems the
concept of {\em commitments} has been used for some time~\cite{commit1,agents}. It has been
suggested that promises and commitments are the same. However, we
shall show that this is not the case. More
seriously, the sense in which the term commitment is used in such
discussions is more stylized than purposely considered and can only
benefit from the discussion in this paper.

\subsection{Promises subject to product/process ambiguity?}
Issuing a promise is an event, and one may say that $A$ promised when $A$ issued
a promise. More common is to view a promise as the outcome, that is a description
of what has been promised, rather than as an event. Nevertheless to some degree
the notion of a promise is subject to product/process ambiguity as described in~\cite{Goddu2011}
and further references cited in that paper. Although~\cite{Goddu2011} discusses the notion of an
argument, its conclusions may hold in the case of promising as well. In particular 
the result of issuing a promise
may be the awareness that a promise has been issued, which is rather 
remote form a conventional understanding of ``product''. The ambiguity is perhaps even deeper
because after having been issued a promise creates a process of credibility and trust assessment and
various other reasoning processes, of which a textual description of ``what has been promised'' constitutes
merely an abstraction. If awareness is understood as a process rather than as a state the matter
becomes quite comparable to the case of an argument.

\section{Modeling the structure of a promise}

Consider the following intuitive idea of what a promise might be:
{\em A promise is an announcement of fact or behaviour by a promiser
to a promisee, observed by a number of witnesses (referred to as the
scope of the promise), whose outcome has yet to be assessed.}

The promiser and promisee are both assumed to be `agents', i.e. humans
or inanimate objects to which we attach identity in the story of promises.
This general description fits the examples that we have already given
and gives some clues as to the constitution of a promise, but it also
opens up a number of questions that need answering.  Already we can
see that this apparently basic definition rests on a number of
assumptions: that we can observe the outcomes of behaviours and that
the outcome of a promise is clear at some single moment of time in the
future, to be measured and verified by an observer. A full account of
this might include a theory of measurement, but we wish to avoid this
level of detail as it binds us to too many details that have nothing
directly to do with the issue. Let us instead try to understand
to essential characteristics for promises and consider what
distinguishes a promise from related matters, such as obligations, 
commitments and other terms.

The model world in which we formulate promises must have the following
characteristics.

\begin{itemize}
\item There must be agents in order for promises to exist.
\item There must be a promiser (or source agent).
\item There must be a promisee (or recipient agent) which might be the same as the source.
\item There must be a body which describes the nature of the promise.
\end{itemize}
We might summarize these attributes with a notation as in~\cite{burgessdsom2005}:
\beq
promiser \promise{body} promisee
\eeq
\begin{itemize}
\item We can leave the body unspecified, but it
must consist of a quality (a type, topic or subject for the
promise) and a quantifier (which indicates how much of the realm of
possibility for that subject is being promised). 
For example: promise quality: ``travel to work'', promise quantity
``on Monday and Friday each week''.
\end{itemize}
Finally, what is implicit in the above is that a promise requires the
transmission of a message, or at least documentation in some kind of
physical form, e.g. a speech act, or a written statement, else it
cannot be made known to anyone except the promiser.  A promise must
therefore have documentation that is made intentionally or
otherwise. 

What then is a promise before we write it down? We shall refer to this
as a {\em possible intention}.  An intention is the basic formulation of a
course of behaviour, which is made internally by (or on behalf of) an
agent.  When an intention is made public, it becomes a promise.  If an
intention is documented or leaked in some way then anyone has a right
to assume it is a promise.

We take it as given that there has to be a source for every promise. A
promiser does not have to reveal its identity of course, so witnesses
to the promise might not know its source e.g. consider the anonymous
threat.  There is no reason to deny the existence of a source
however.  The lack of such information about a promiser is simply a defect in
the knowledge of the receiving agent, but one would normally prefer to
assume a consistent picture of promises and infer the existence of an
anonymous promiser. This justifies our postulating the source.

\subsection{Promises as documented apparent intentions}

Stoljar~\cite{stoljar1} writes that
a promise is an announcement of an intention. and he argues that any offer
must be a promise as well. Indeed a key characteristic of a 
promise is that it documents an intention, so
let us explore the idea of intentions in more detail.  Intentions turn out
to be a lowest common denominator for all of the concepts discussed in this paper
and thus have a special importance. Unfortunately from a document one cannot
conclude to what extent the intention expressed in it is real. That indeterminacy
gives rise to an additional complexity of our definitions.

Since promises involve communication we require a notion of the spread
of information amongst the agents. We use the term {\em scope} (as
used in computer science) for this.  A scope is simply defined to be a
set of agents.  For example, the scope of a promise would typically
refer to the promiser and a list of witnesses to the promise,
e.g. those who heard to utterance or those who saw the written
document.

\subsection{Intentions}

In the realm of all possible formulations about agent behaviour 
the concept of an intention stands out as an important foundation.

\begin{definition}[Current intention of an agent $A$]
A current intention of an agent $A$ is description of a possible behaviour, 
or goal, or objective, or state of affairs,
that is contemplated by $A$ with the understanding that it can be and 
preferably (for $A$) will  be brought to realization.
\end{definition}

\begin{definition}[Possible intention for an agent $A$]
A possible intention for an agent $A$ is a description of a possible behaviour, 
or goal, or objective, or state of affairs,
that may but need not currently (at the time of qualifying the description) be 
contemplated or preferably brought to realization
by $A$, and which might be in some (possibly different) 
circumstances a current intention of $A$.
\end{definition}

Obviously a current intention is also a possible intention. 
But if an utterance of $A$ announces a possible intention that
need not be a current intention, it may only appear to be a current intention. 

The components of an intention are as follows: a source agent who formulates the
intention, a target agent if the intention is directed at a potential
subject, and a body which explains the quality and quantity of the
intention. Only the source of an intention
knows about the intention, i.e. the scope of an intention is the source only.
There are no witnesses.

Now we must be careful: the set of all possible intentions should be distinguished from actual
instances of intentions selected by an particular agent. We shall
sometimes use the phrase ``possible intentions'' to mean this full set
of abstract entities to emphasize when we wish to signify a general
description of behaviour rather than an individual
agent's decision.

\subsection{Commitments}

To commit to something is to make a decision in favour of it.  The
issue of commitments is therefore about the favourization of
intentions. Commitment is a personal decision and has nothing to do
with physical representation or communication, thus the issue of
commitment precedes any discussion of promises.  A commitment has a source, a
target and a body, i.e. it is an intention. Like an intention it
has no physical representation and does not have a non-trivial scope.

Once an intention becomes a commitment we often assume that some point
of no return has been passed in the act of committing (deciding) about
the particular intention.  i.e. adding the intention to a list of
commitments.  For example, in a game of chicken in which two cars
drive towards each other to see which one will swerve off first, a
driver has committed to not swerving when the decision to not back
down has been made~\cite{schelling1}. This might have certain
irreversible consequences, but it is difficult to generalize the idea
of irreversibility in examples of this kind. What commitment
essentially boils down to is the elevation of some intention beyond an
arbitrary threshold. In other words, in the universe of intentions
there is a subset of these which we may call commitments. 

\begin{definition}[Commitment]
Commitments are current intentions that we are committed to. We may call them
intended intentions, or equivalently real intentions, intentions that
we hold, or committed intentions. The commitment of an intention exceeds its 
merely being current
in that it is stable and persists in time until some achievement of the intention
will take place or until some overruling considerations invalidate the commitment.
\end{definition}

When passing by a shop one may feel a current intention to buy a nice gadget, only
to be relieved of that intention (or rather its currency) after noticing its price. If however,
the price is quite good, but the shop is closed at the time of passing along, then a current intention
to buy the same item can become activated with the status of a commitment, only to be 
terminated when the item has been acquired or when unexpected problems turn out to stand in the way.

\subsection{Expressing intentions}
Making intentions known to other agents is the essence (meta-intention) of promising. 
We will develop some
terminology for such acts of expression.
An intention is not necessarily announced by the agent holding it to any other agents. 
Indeed, we may now \emph{define} any intention
that is announced to be a promise. Conversely  we notice that any promise that has
not been announced is merely a possible intention.
Some intentions are desirable while others are absolutely undesirable
and an agent might never choose them, yet they are possible intentions
nevertheless. The fact that such behaviours can be intended
is enough for them to qualify as possible intentions.

Intentions must always be thought of as belonging to a specific agent. Those
intentions which are actual plans of the agent are called its commitments. Other
phrases for a commitment that we may use are: intended intention, or real intention.

Due to the static nature of our account we pay no attention to the
process by which an intention might become a commitment or vice versa.

\begin{definition}[Intention utterance]
An agent $a$ produces an intention utterance if $A$ produces an 
expression of a description of a possible intention.
\end{definition}

What matters for our discussion on promises is intention utterances that seem to be real.
That leads to the idea of an apparent intention utterance.

\begin{definition}[Apparent Intention utterance]
An utterance expressing a possible intention (of a principal agent) with the 
contextual appearance of an intention. Apparent intentions, 
may be drawn from the following range:
\begin{description}
\item [\em Real intention:] (alternatively: commitment, true intention, or intended intention) what is announced 
corresponds to wha the agent expects that will happen, or that (s)he will do,
or what holds or what will hold. 

In other words the apparent intention is real if it is a commitment (and therefore current).

\item [\em Incidental intention:] (alternatively: non-committing current intention) what is announced 
corresponds to why the agent expects that will happen, or that (s)he will do,
or what holds or what will hold, but only a the time of expression. 

\item [\em Indifferent intention:] (alternatively: quasi-intention) the issuer has 
no current intention corresponding to the
utterance, and no current conflicting intention either. 

An indifferent intention
is currently contemplated as a possible behaviour, goal, objective, or state of affairs, but its bringing about
is not preferred, and thus an indifferent intention is not a current intention.

\item [\em Deceptive intention:] (also: misleading intention) the 
announcement might seem to be real for an audience in 
scope but it is a lie from the perspective of the promiser.

A deceptive intention is incompatible regarding realization with a current intention, though this may be only
known to the principal agent.

\item [\em Invalid intention:] (alternatively: manifest lie) all observers may notice a discrepancy 
between what is announced and the facts.

The invalidity of an invalid intention will become clear to agents in scope of that utterance.
\end{description}
\end{definition} 

The idea of an apparent intention is that at face value it is like an intention from the perspective of an
observer but there is a considerable degree of freedom in connection with a so-called underlying intention, 
the existence of which we postulate in the following definition.

\begin{definition}[Underlying intention (of an apparent intention utterance)]
Given an apparent intention utterance of an agent, there is an underlying 
intention (which need not be comprised in the same utterance) as well. We will distinguish five cases, 
corresponding to the case distinction
of intention utterances:
\begin{description}
\item [Real intention:]  The underlying intention of a real intention is that same intention.
\item [Incidental intention:]  The underlying intention of an incidental intention is that 
same intention which is known to be consistent but is non-committing as well.
\item [Indifferent intention:] The underlying intention of an indifferent intention is empty.
\item [Deceptive intention:] The underlying intention of a deceptive 
intention differs significantly from the (deceptive) intention.
\item [Invalid intention:] The underlying intention of an invalid 
intention differs noticeably (for observing agents) from the (invalid) intention.
\end{description}

\end{definition}

We will assume the agents keep underlying intentions private. 
Otherwise new levels of complexity emerge as underlying intentions may 
turn out to split over the same distinction of four cases recursively.

\subsection{Promises}

A promise is the physical publication of an intention within a certain
scope. This suggests that there must be some agent to observe the
promise and its outcome which in turn requires the expensive notion of a
theory of observation so we shall tackle this issue separately (see
section \ref{assessment}).

\begin{definition}[Promise]
A promise is an apparent  intention  of an agent, 
(the promiser or promising agent) 
the utterance of which has been documented 
within a scope that goes beyond the promiser.

According to the definition of intention utterances, a promise brings
with it an apparent intention and an underlying intention, and five cases can be 
distinguished for promises: real, incidental, indifferent, deceptive, and invalid.
\end{definition}

Promises thus have scope. Formally intentions also have a scope, but
the scope of an intention held by an agent is by definition limited to
the agent (source) itself. An intention could be leaked deliberately
(e.g. to the press, in order to influence someone). This might be a
form of leverage, or an attempt to impose an obligation on some party
in scope. However, at the instant an intention expands in scope to
encompass more agents it becomes a promise. A so-called letter of
intent, for example, is a promise rather than merely an intention.

The time aspect of promises presents further challenges.  Intentions
can become outdated by events.  An event which is found to fulfill an
intention documents the implicit promise, since one must admit to the
intention in a wider scope.  Conversely, the documentation for a
promise does not have to last for ever; if documentation of a promise
disappears completely, it reverts to being an intention.  A promise to
oneself is in excess of a mere intention because  it has been documented.

The distinction between the promisee and the scope of the promise is key
to understanding promises.  Suppose someone intends to arrange a
surprise birthday party for their friend. Initially this is an
intention. The intention is written in a diary or mentioned to a third
party and it becomes a promise. The target is not in the scope of the
promise however, so the promise remains unknown to the jubilant.
However, suppose that before telling anyone else, the promiser
destroys all evidence of the promise by tearing out the page of the
diary, effectively withdrawing knowledge of the promise, then the
promise reverts to being just an intention. But as long as knowledge
of the promise remains ``out there'' in the world, it remains a
promise that has been made.

\begin{definition}[Keeping a promise (relative to an observing agent)]
A promise is kept, for the perspective of an agent $A$ (e.g. the promisee), 
if an action is performed or a state of affairs is reached
that complies, according to $A$'s assessment or observation, with the body of the promise. 
It is plausible but not strictly needed to require that
once a promise is kept some causal relation with promiser behaviour can be found.
\end{definition}
Although in a static theory of promises the keeping of promises is not center-stage,
having some some characterization of it is helpful. Critical to the definition of keeping a promise 
is relativity w.r.t. the observing agent. If a promiser assesses a promise $p$ as having been 
kept while a promisee $B$,
perhaps supported by other agents in scope of the promise doesn't agree, 
that state of affairs casts no
doubt on the status of the promise $p$ as such. On the contrary this situation may 
have been precisely what
$A$ had in mind when issuing promise $p$.

\begin{definition}[Keeping a promise (in absolute terms)]
A promise is kept in absolute terms, if it is kept according to all agents in its scope.
\end{definition}

\begin{definition}[Broken promise]
A promise is broken if after it has been issued a state is reached at which the promise has not
been kept, from any conceivable (though reasonable) perspective, 
and from which it will certainly not be kept anymore either, again from any conceivable (though reasonable)
perspective that an agent in its scope mint have.
\end{definition}
We will now discuss the relation between promises and obligation in some detail.

\begin{definition}[Promissory obligation]
With each promise  of an agent $A$ an obligation is connected, the so-called 
promissory obligation. It is that obligation to which the agent has become
obliged by making the promise.
\end{definition}
Promissory obligations are an important tenet of the philosophy of promises, and
we do not deny their existence. This definition allows significant freedom. It can be read as follows:
a promise theory will allow for a notion of a promissory obligation. That notion may be doe 
weak that it provides empty promissory obligations only. It is not assumed that a promissory
obligation explains a promise in terms of a philosophically grounded concept of an obligation.
However, we will oppose to what we call
obligationism.

\begin{definition}[Obligationism]
With obligationism we denote the viewpoint that (i) promises are characterized by
a unique capacity to (auto)generate an obligation (specifically the promissory obligation) 
for the promising agent, and that (ii) the essence or content of a promise is fully captured by its 
promissory obligation.
\end{definition}
An argument in favor of obligationism is found if one assumes that the promissory obligation
captures the essence of the obligation so that the concept of a promise can be reduced to that of
an obligation. Obligations being known from deontic logic and form various strands of philosophy,
one may consider this reduction to constitute a worthwhile simplification rendering the construct
of a promise redundant. We don't support this argument.

\begin{definition}[Non-obligationism]
With non-obligationism we denote the belief that obligationism is false.
\end{definition}

\begin{definition}[Strong non-obligationism]
With strong non-obligationism we denote the belief that obligationism is false and that in addition
the concept of promise may be accounted for without making use of the concept of an obligation.
\end{definition}
Preferring non-obligationism over obligationism, and convinced of the relevance of
strong non-obligationism, we will propose a strongly 
non-obligationist conception of promises. 
Arguments for (strong)
non-obligationism have been detailed in \cite{burgessdsom2005}. 
\begin{definition}[Restricted strong non-obligationism]
With restricted non-obligationism we denote the belief that obligationism is false, 
and that for a large class of promises, though not for all, the concept of promise may be 
accounted for without making use of the concept of an obligation.
\end{definition}
If in court a witness promises to tell the truth and nothing but the truth, an obligation is created. 
Remarkably that obligation can only be created by means of a promise. Restricted strong
non-obligationism does not claim that this promise, though being prior to its promissory
obligation, can be fully understood without contemplating the concept of an obligation.

A recent 
argument can be found in \cite{BergstraL2013} where payments in a peer-to-peer
system for financial transfer an store of value are considered autonomous actions
to the extent that such payments cannot be obliged by definition. Payments can be promised 
however, with the corresponding promissory obligation constituting no more than 
acquiring some certainty that the payment could be effected if the agent wishes
to do so. 

\subsection{Deceptions -- non-intended promises}
Although we have distinguished five cases for promises, commitments stand out
as most important immediately followed by deceptions, with the other classes 
entering the picture in order to have a complete story. We add some further comments on deceptions.

Understanding deceptions (or lies) is also an important step in
clarifying the relationship between intentions and promises, because
it is possible for an agent to have two different intentions in play
at the same time: a commitment and an announced intention (i.e. a
promise) which are not compatible. Incompatibility means that striving
for both intentions simultaneously is fruitless because their
realizations cannot be combined.

In a deception, there is always a source and always a target and the
target cannot be the same as the source, as an agent cannot
(intentionally) deceive itself. Furthermore, we maintain that the target of a deception
must be in scope, so there must be a physical documentation and hence
a deception necessarily involves a promise and not merely an intention.

\begin{definition}[Deception]
A deception consists of two intentions: a documented intention (i.e. a promise) and
a non-documented commitment, which are incompatible. 
\end{definition}
The non-documented commitment will also be called the hidden intention.

In a deception the hidden intention is more important than the
witnessed one one and we might refer to it as the dominant
intention. This simply expresses that it is a commitment while the
promise contains merely a ``possible intention''. It is the {\em real}
intention of the agent (``intended intention''), while the intention
in the promise can merely be described as {\em non-real}.  If the
dominant intention should be rescinded, a deception will revert to
being a promise, but this is only known to the source.

\subsection{Positive promises}

A deception is the augmentation of a promise with an {\em incompatible}
intention. We should like a name for the augmentation of a promise
with a positive intention.  We might call this a positively intended promise, 
or more briefly a positive promise.

From these slightly strained terms, we can now appreciate why the
concept of a promise is in fact so important. A promise is simply a
promise (the documentation of an intention), regardless of what lies
behind it.  Any internal priorities or considerations are hidden from
the view of other agents and cannot be observed.  Thus, promises are
an independently important concept because we can (indeed must) talk
about promises without discussing the basis on which they are made.

When a promise is issued, we are neither required nor able to confront
the truth or falsity of the promise.  Indeed, as soon as we ask such
questions, new issues such as trust and a plethora of other subjective
issues come into play.  Such issues are probably un-resolvable in a
logical sense. However, what we assume is here that no matter how
trustworthy a promise might be, it can increase or decrease our
certainty of a promised outcome and thus it bears an {\em influence}.

\subsection{Promises as specifications in technology}
Suppose a user community takes an interest in machines $M$ that 
compute a function $f$ from a finite domain
$D$ to a finite range $R$, with the only constraint that for each $d \in D$, 
the result $r=f(d)$ satisfies some criterion $\phi(r,d)$.  
Thus for $f$ we have an implicit definition that is likely to be satisfied by many different implementations.
A provider $P_m$ of these machines may deliver the good (say machine $M$) in a 
box equipped with the following promises: (i)
the expected time of computing $f(d)$ is below $E_t(d)$, (ii) the distribution of running times is normal, 
(iii) the spread is $s$, and
(iv) energy consumption when running is below $e$ (per unit of time).

Are these four assertions about $M$ best viewed as promises, or is computer science able to 
provide a better terminology in this case? 
We first notice that it is certainly  conceivable that the setting is so complex
that it is implausible that any assertions about $M$ like the 
mentioned promises can be proven about it as correctness
assertions.  This may be the case 
in spite of the fact that each of these assertions are in essence well-defined 
mathematical, logical, or  combinatorial, statements. 
No amount of testing can compensate for that difficulty. 
The subjective status of such promises when understood 
as propositions or assertions about the behavior of $M$ will not go away simply by not 
thinking about them in terms of promises.

A very important flexibility results from viewing these assertions as claims about $M$ that have been
cast as promises. In order to appreciate this flexibility we will make
the plausible assumption that the provider of $M$ has arrived at those claims 
after due experimentation, 
thus giving them initially a status comparable to that of a scientific hypothesis, at least from the 
provider's perspective. 

We consider the circumstance that the user community of machines of type $M$ 
and of similar competing equipment that has been
provided by competing providers in combination with different packages of promises feels 
dissatisfied with $M$ in the light of $P_m$'s promises. 
Now the members of the user community have the option to update their trust in $P_m$, rather than to
contemplate modifications of the promises that $P$ has made. 

It should be noticed that, apart from forgetting entirely about $P_m$'s promises 
regarding $M$, the only alternative way to accommodate their dissatisfaction that is open for
the members of the user community, 
is to threat these promises as scientific hypotheses which must 
now be updated in a systematic fashion on the
basis of new information gained during usage. 
But following that systematic path of renewed hypothesis formation is impossible without being given
full information on how $P$ arrived at these opinions in the first place. And such information is likely
to be unavailable to the members of $M$'s user community.

These remarks serve as an illustration of the principle mentioned in  Paragraph  \ref{Reduction-of-uncertainty} above.

\subsection{A technical refinement of the example}
The setting just sketched may seem artificial, but with the following example we 
will demonstrate that it makes practical sense. Let $2^{n}$ represent
the set of bit sequences of length $n$. SHA-256-512 is the secure hash algorithm SHA-256 applied to 
$2^{512}$ (see~\cite{FIPS2008} for an authoritative description and \cite{BergstraM2013} for an alternative 
formalization). We will abbreviate 
SHA-256-512 to $h$. As a domain $D$ we choose
$D = 2^{256} \times 2^{256}$, and the range $R$ is $2^{256}$. The criterion $\phi(d) = \phi(d_1,d_2)$ works as follows:
\[\phi(d_1,d_2,r) \Leftrightarrow \texttt{bs2n}(h(d_1 \mid \mid r)) \leq \texttt{bs2n}(d_2). \]

Here $\mid\mid$ denotes string concatenation and $\texttt{bs2n}$ converts elements (bit sequences) of $2^{256}$ interpreted
as naturals in binary notation with big endian representation to  
natural numbers.\footnote{%
Ignoring any knowledge of SHA-256-512 and assuming that it works randomly an estimate of the probability that
some $r$ satisfies the mentioned condition is $\texttt{bs2n}(d_1) \cdot 2^{-256}$. This provides an expectation
of the number of inputs that must be tried before a bit sequence $r$ that ``solves the combinatorial problem'' is found.}

ASIC based machines that compute functions $f$ according to these concrete specifications, 
modulo some details, are now on the market as so-called 
Bitcoin (see \cite{Nakamoto2008}) mining tools.\footnote{%
Such equipment is being sold for instance under the trademarks Avalon and Butterfly.} 
We claim that
viewing claims about these machines put forward by their providers as promises is convincing. 

SHA-256-512 is presented by FIPS (see \cite{FIPS2008}) as a secure hashing function 
which is for that reason supposed to be 
resistant against so-called first 
pre-image attacks: given $r \in 2^{256}$ it is very difficult to find a string 
$d \in 2^{512}$ with $h(d) = r$. Although 
expressing the algorithmic difficulty of finding $d$ given $r$ as a property of SHA-256-512  is a 
mathematical issue in principle, no indication of a proof of the difficulty of that problem has been found. 
Nevertheless, the assertion about $h$ that the mentioned problem is computationally difficult to solve 
can be qualified as a 
scientific hypothesis maintained by the computer science community at large rather 
than as a promise made by FIPS.

Nakamoto's design of Bitcoin turns SHA-256-512 into a parametrized problem 
(the Bitcoin mining problem: computing $f(d_1, d_2)$,
necessarily fast in order to win a competition and preferably at low energy costs on cheap but dedicated 
machines in order to make a profit as well) 
which is also assumed to be difficult to solve, and which is assumed to grow 
exponentially more difficult when the number of leading zeroes of $d_2$ increases. 
Otherwise
Bitcoin mining will eventually be cracked so to say. In the absence of a 
thorough scientific analysis of the Bitcoin mining problem
the difficulty of that problem is probably adequately viewed as a promise 
made by Nakamoto and reproduced by the open source 
developer community around Bitcoin.

\subsection{About the philosophical literature on promises}
Having worked out a view on promises in this Section, we are in the position to make some further 
remarks concerning the connection of non-obligationist promissory theory as advocated above and the
plurality of views on promise that has been developed in philosophy. We will use the illuminating entry on 
promises in the Stanford Encyclopedia on Philosophy as a guide (see~\cite{Habib2008}). We refer to that 
survey paper only and don't repeat references cited there.

Our non-obligationist
promise theory (NOPT) takes accounting for promissory obligations and measuring their
force in comparison to other kinds of obligations for an issue of secondary pragmatic importance only.

NOPT may be considered a branch of act-consequentialism, and also as a brach of the expectational 
theory of promises.
In mechanical terms NOPT is connected to speech act theory and 
promises may be considerd a sublcass of Searle's
commissions, and of the illocutions of Austin. 
Expectationalism portrays a view of promises not unlike that of NOPT,
where trust and modifications of trust are the key parameters against 
which promises and promise keeping are measured.
NOPT seems to differ from expectationalist views in not paying attention to whether or not failing to live up to
a promise and thus failing to meet an expectation created by and trusted in connection with a promise is wrong. 
Rather
than  thinking in terms of wrongdoings the future impact of not keeping a promise on the interaction 
between promising and
trust maintenance and its secondary effect on the calculation of expectations from a promisee's perspective is
considered central in NOPT.

NOPT has been worked out in such a way as to be relevant for automated,
or artificial, agent communities as well as for communities of human
agents. For that reason NOPT has a bias, at least in principle, towards
(i) quantitative aspects of the creation and maintenance of expectation,
(ii) logical aspects (informal logic) of reasoning on the basis of expectation and trust,
(iii) trust maintenance and the interaction between trust and reputation dynamics,
(iv) temporal aspects of a promise life-cycle,
(v) interaction between different promises.

The above focus of NOPT is at cost of a diminished focus on the following aspects:
(i) moral aspects and ethical valuation of actions,
(ii) the special status of promissory obligations (which seem to be rather useless in a world of artificial agents),
(iii) the ramification of speech acts (though NOPT must view a promise as an act of communication),
(iv) promises in relation to contracts (for which the importance of obligations is not denied), and with various
forms of contractualism and contractarian theory,
(v) relations with Law.

\subsubsection{Promises in the philosophy of Law}
There is a large body of literature in the philosophy of Law about the role of promises in general, and in the context
of contracts in particular. In~\cite{Pound1959} the question why and to what extent 
keeping a promise ought to be legally 
enforced is discussed and seven different views on the matter are distinguished. 
What matters for us is that the author takes promising as a given institution and 
understands the conception of legal instruments as a variable, rather than the other way around. However,
thoughout this paper promises invariably give rise to expectations, whereas the status of promissory
obligations is much less obvious, though such obligations if any are always connected to promissory
expectations, which conforms to our own viewpoints.

In~\cite{Craswell1989} the case is made that philosophical explanations of why promises are bining
are less critical to the development of contract law than many authors seem to defend. 
Instead a notion of efficiency
is brought forward which might have greater explanatory power reading the role of promising.

\subsubsection{Promising in anthropology}
in~\cite{KornK1983} the observation is documented that in some communities promising does not exist
because promises are not considered as creating (binding) promissory obligations, which is taken to be
an essential aspect of the institution of promising. We might turn the message of the paper around and draw the conclusion that there exists anthropologic evidence for the feasibility of human communities with a non-obligationist
conception of the institution of promising.
\section{Promise valuation, assessment, reasoning, and logic}
In this section we discuss that  a promise may have tome value 
to its promiser as well as to its promisee. Further the strength of a promise is 
on display as well as as variety of assessment methods and finally an
embedding of the tropic in ``Informal Logic'' is given.
\subsection{Promise value}
Promises are valuable to agents, because they help to deal with uncertainty.
Because certainty is key, a promise is worth nothing unless there is
trust. Zero trust makes promises worthless. Trust might be based on a
history of keeping promises or, in our terminology, on a history of
positive assessments about a succession of promises. So there is
a symmetry between trust and promises that must be broken to solve the
dilemma.

If there is trust, a promise about future behaviour does not need to
be perceived as an obligation on the promiser but as an indication
that best effort will be respected.  If a given agent $X$ does not
trust the promiser however, it might assume that the promise implies
an obligation on the promiser. This perception of obligation by $X$
does not of course imply an obligation perceived by the promiser.
There is a fundamental subjectivity in these perceptions.

The value of a promise to any agent in its scope is an expectation of the eventual
benefit. Suppose, then that $A$ promises $B$ 400 dollars per year. $B$
promises to wash $A$'s windows at this price.  Both are satisfied with
the value they get from this arrangement and prefer not to question it
too much as this could unleash all kinds of consequences.  Observer
$C$ can see that the values are quite mismatched, or that $A$ is
getting a poor deal by its judgement, but $C$ also cannot deny that
the relationship is stable because both $A$ and $B$ are happy.

\subsection{Promise strength}
The above example illustrates that once promises are made and have 
been noticed by an audience consisting of
agents in its scope, these agents engage in a process of trust management. 
The strength, or force, of a promise, viewed 
from the perspective of an agent in its scope, quantifies the expectation that it will be kept. 
The same promise may have different strength for different members of its audience, and the 
evolution of that strength may differ 
between agents with different experiences. Promise strength combines 
(i) a promisee's promise independent trust in the promiser, 
(ii) the credibility of the promise (given the promiser), 
and  (iii) the promise type related trust of the promisee in the promiser.

It must be stressed, however, that the promiser, the promisee, 
and all agents in scope are individually responsible
for their management of the strength they attribute to a promise. 
Aggregate trust management may be used, but is is
necessarily a voluntary matter for all agents involved.

\subsection{Assessments}\label{assessment}

The notion of whether promises are kept or not is central to their
sustained usefulness in any application domain, thus we need to make mention of how
this comes about in a theory of promises. It would be possible to go
overboard and delve into the complexities of observation and
measurement tin an attempt to provide a satisfactory answer but that is not in the
spirit of this paper. We seek instead a simpler notion which is at the same
level of abstraction as the concepts of promise and intention that we
have introduced thus far. We call this the concept of {\em
assessment}.

\begin{definition}[Assessment]
An assessment is a subjective statement made by an agent about whether
the intentions of itself or of another agent were fulfilled. 
\end{definition}
Our notion of an assessment is more generic and less quantitative than
a verification. It is both subjective and not {\em a priori}
linked to observation. In a static theory of promises and intentions the existence
of intentions as well as the value of assessments is linked to state parameters like time.
Thus, for an intention of agent $A$, in existence at time $t$, and known to be by way
of a promise that $A$ had issued before $t$, it may be the case that agent $B$'s 
assessment, made at time $t'>t$ is positive (or negative).

At this level of description, we need not say any more about it than
this.  What is important is how a witness to the promise
assesses the fate of the promise. Such an assessor may or may not
feel obliged to assess a resulting state or activity in a particular way, might promise
to conform to certain criteria, and so on. What matters is only the
assessment, which might or might not be rationally obtained.  We
believe that this is a fair model of the world in which we live.

Assessment involves a variety of possible routes to
inference, i.e. there are different kinds of assessment. This is a
subjective issue, but this should not be viewed as a weakness of our
theory, it is an important feature of the subjective nature of individual
assessment.

\subsection{Promise related reasoning processes}
Promise descriptions are symbolic representations of actions, and events and states of affairs
that allow systematic transformation processes. Transforming the physical representation of
a symbolic description of a promise might be considered an instance of  reasoning. 

The logical complexity of promises becomes apparent by taking into account  the
diversity of reasoning processes related to a single promise that are
performed by promiser, promisee, and agents in scope of that promise. 

The following reasoning processes can be distinguished in relation to 
a single promise
$p$ with promiser $A$, body $X$, promisee $B$ and with agent $C$ in its scope 
($C$ may be equal to $B$). We assume that some representation of $P$ and its components,
or of components in a stage of preparation of $P$ is
available to an agent that performs reasoning about it. No analysis is suggested of the means
of representation used, but this range will include mental representation as it occurs within an
animate agent and also formal representation amenable for some formal reasoning process
as it may be implemented within an inanimate intelligent agent.
\begin{description}
\item [Promise preparation:] a reasoning process performed by $A$ leading to the design, 
timing and issuing of $p$ by $A$.
\item [Credibility analysis:]  a reasoning process where agents $C$ in scope of $p$ determine 
the credibility they assign to $A$'s promising $p$ in the light of general facts known of $A$ 
(but excluding specific historical information about the individual behavior 
of $A$ as a member of its agent class.)
\item [Prior trust determination:] reasoning processes performed by $C$ (provided that $C$ is in 
the scope of $p$) aimed at
(i) determining the trust $C$ had in $A$ before becoming aware of $p$ 
($C$'s prior trust in $A$), and (ii)
specifying which expectations are generated by taking notice of $p$.
\item [Counter-promise deliberation:] reasoning processes performed by $B$ concerning plausible 
counter-promises that may be
issued in by $B$ reply of taking notice of $p$ (provided $B$ is in scope of $p$).
\item [Promise impact prediction:]
(this may be done with the understanding that $B$ has issued one or more
plausible counter-promises) (i) reasoning processes performed by $B$ (when in scope of $p$)
and by $C$ (any agent in scope of $p$) aimed at determining the (change of) 
expectation (as viewed by $B$ resp. $C$) 
that $p$ creates in $B$ (and that $A$ intended to generate), (ii) and reasoning processes
aimed at modification of plans (held by $B$ or by $C$) given the 
change of expectations held by each of them that was brought about by taking notice of $p$.
\item [Promise assessment:] reasoning processes performed by $C$ concerning (i) the way in 
which $C$ will assess whether
or not $A$'s promise is kept, (ii) assessing the latter by means of the 
preferred assessment method.
\item [Promise withdrawal monitoring:] $A$ may at some later stage issue another promise, say 
$q$, the keeping of which is not compatible with keeping $p$. It that case $q$ qualifies as a withdrawal 
of $p$.
An agent $C$ applies a reasoning process that monitors and evaluates subsequent promises issued by 
$A$ to determine whether these may be viewed as breaking $p$ to an extent that $p$ 
may be considered having been withdrawn. 
\item [Trust updating:] reasoning processes in place for each agent $C$ in scope of $p$. 
The process aims at updating $C$'s prior trust in $A$ 
in accordance with the result of the assessment  $C$ makes 
concerning the degree of keeping $p$ that is achieved by $A$.
\item [Reputation processing:]
a reasoning processes performed by each agent $C$ in scope of $p$ aimed at exchanging the effects
of trust updates between different agents.
Reputation flow permits an agent $C$ while having no prior trust assessment of an agent $A$ 
to acquire an initial trust of $A$ by taking notice of evidence gathered about $A$ by other agents.
\end{description}

\subsection{Promise related informal logic}
Connections between promises and informal logic are of a diverse nature. 
Such connections admit a general classification. Here are four forms of such connections that
may be imagined, the first three of which may be considered relevant in the case of promises: 
\begin{description}
\item [Promises as a contribution to informal logic:] Promises considered an additional
entry in the catalogue of entities that are central for informal logic.
\item [Viewpoints from informal logic supporting promise theory:] Definition and demarcation of the 
concept of a promise profiting from existing developments in informal logic.
\item [Methods of informal logic applied to promise dynamics:] Promise related reasoning 
mechanisms have in common that a strict  formalization is implausible. 
A significant variation between promise related reasoning patterns for 
individual agents is plausible. This variation is a feature of informal logic rather than 
of formalized logic.
\item [Promises as an alternative to (parts of) informal logic:] We are not aware of parts of 
informal logic that would profit from being rewritten in the language of promises.
\end{description}

\subsubsection{Specific links between promises and informal logic}
 Concerning these three forms of connections that matter in the case of promises 
the following more specific 
remarks can be made:
\begin{enumerate}
\item Promises are utterances like remarks, questions, announcements, replies, and 
obligations.\footnote{%
While an obligation is not a speech act  and not the result of a speech act 
a promise is likely to be the result of a speech act. The dynamos of obligations and promises differs.
Such differences an be found with utterance types remark, question, announcement, reply, 
and (to a lesser extent) prediction as well.}
Special of promises is the dynamic interplay with credibility and trust. 
Issuing a promise differs from producing a statement 
in  discourse (dialectical use), or from making an argument (alethic use), 
or from rhetorical language
(though rhetoric may involve promises and deceptive promises). 
We mention \cite{Pinto2009} for an application of this distinction. In that paper reason for action 
are studied in detail, and reasons for issuing a promise can be considered a subtopic within that theme
worth of independent investigation.

\item Credibility assessment may involve reasoning by analogy (similar agents have kept similar
promises in similar circumstances and therefore there promise is credible), 
and arguments from authority (another highly regarded agent claims that the promise is credible). 
Pattern matching techniques may be 
used to support an agent in finding similarities that allow for comparative assessment of credibility.
The form of reasoning is similar to ``story similarity'' as discussed in~\cite{Walton2012}.
\item Trust assessment may involve arguments from authority and as well as various 
forms of probabilistic and plausible reasoning. We mention~\cite{Stephens2011} which 
deals with systematic reasoning
given the absence of evidence, a very plausible condition for agents who need to 
assess trust in a promising agent. Many examples of trust based reasoning are given in~\cite{Pinto2013}.
Although that paper focuses on trust as a human behavioral or mental feature 
most examples make sense in a context of artificial examples just as well. 
In~\cite{Walton2006} one finds an analysis of detecting
and avoiding circular reasoning when establishing or updating trust.
\item Both trust and credibility are conveyed and created through a 
combination of network based reputation development, 
which is an instance of distributed reasoning from authority, and distributed
plausible reasoning.
\end{enumerate}

\subsubsection{Nonlogical processing of promise bodies}
Agents aware of promises that have been issued will perform processing on the basis of 
their own local 
representations of promise statements. From a symbolic perspective this
involves symbolic processing on (representations of) promise types and promise bodies.
 
But there seems to be no need for a notion of validity for the processing of promise statement
representations that is performed by agents in scope of a promise. Whether or not promising
works out well to the extent that it is of sustained use in some context is a learning outcome 
by agents who may act as promisers and promisees.
 
Besides alethic use, dialectical use, and rhetorical use, a behavioral use is conceivable. 
Behavioral use of promises does away with a principled approach to the
 promise types and the meaning of promise bodies as well as with valid 
 reasoning on the basis of these ingredients  
and admits any form of associative processing by agents involved that suits their objectives.

\section{Promises versus obligations}
The importance  of the notion of promises may be 
questioned by those who consider promises
as a concept secondary to obligations. If one views obligations as the
primary concept, the value or importance of promises unavoidably
shrinks. We shall now survey advantages and disadvantages of
obligations as an alternative cornerstone of a theory of multi-agent systems. The
discussion will be somewhat asymmetric because we will not base our
comparison on a proper definition of the concept of an obligation
(which we cannot fathom). 

Having explained intentions and promises, let us now try to
describe the notion of obligations in the same manner. The intuitive
notion of an obligation seems straightforward, but it proves to have difficult
properties. We refer to Jackson \& Altham \cite{JacksonA1988} for a convincing 
exposition of these complications.

We might try to think of obligations in a straightforward way, for instance:
{\em an obligation is an intention that is perceived to be necessary by an agent}.
This certainly captures some of the characteristics that we understand by
the term, but it also leaves many questions unanswered: is the feeling of the necessity
voluntary or forced, a matter of survival or simply an authoritarian convention?

An obligation, or more precisely the impact of an obligation in general,
as felt by an agent within its scope, falls into the category of possible
intentions. We will speak of an induced intention, the intention being achieving 
compliance with the obligation.
The induced intention must have source, a target and a body, and the body must
have a quality and a quantity. The source and target are now somewhat
difficult to understand however. Unclear is further to what extent an induced intention
is current, and if so if it is a commitment. Can it be indifferent, 
or even a deception or simply invalid? 
Beyond this, we shall not attempt to define obligations more carefully
in this paper. We shall merely state some assumptions about them.

An obligation can be imposed by external conditions, e.g. by the
expected behaviour of external agents, by laws threatening sanctions
etc, or it can be self-imposed by codes of personal behaviour which an
agent holds to be {\em necessary}. But this imposition suggests the
action of a force which attempts to induce a commitment in another
agent (or itself).  An obligation is a possible
intention which may or may not be current and may or may not 
have the status of a commitment.  In
any case the agent is aware of any compelling reasons to include the
intention in the portfolio of commitments, either from within itself
or without due to external forces.

It seems natural then to refer to the source and target of the induced
intention as being the agent in which one attempts to induce the
intention, and the recipient of the intention respectively. However,
the source of the obligation itself might not be an agent at all, but
merely a set of external conditions, norms, experiences or other
information acquired by the agent that lead to a perceived priority.

Note again that even `forced' behaviour can be classified under the
realms of (possible) intentions since all behaviour can be
intended. Again, we emphasize that this does not imply that a coerced
agent holds the intention that is being forced upon it. Nor does it
say anything about whether the agent is able to resist the force or
not, or whether it matters if an obligation is self-imposed or
externally imposed.

Viewed from the perspective of an agent, the notion of an obligation immediately 
seems significantly  more complicated
than an intention or a promise and does not seem to be close to the
notion of either promises or intentions.

We hold that obligations are far from being a reliable tool for
ensuring compliance. If a law-giver wanted to ensure the compliance of
an agent, a better strategy would be to obtain a promise from the
agent, and to convince it to view the intention as a {\em commitment}
since the law-giver could never know whether the agent had indeed
committed to the body of the obligation.


\subsection{Obligation strength}
Obligations can be assigned a measure or degree of strength from the 
perspective of an audience just as promises.
When assessing the strength of an obligation an agent will first of all 
estimate the degree to 
which an obligation applies to its own context, rather than
the degree to which the obliging entity is entitled to do so.

The divergence between promise strength and obligation strength appress strikingly when one
appreciates that, unlike promise strength, obligation strength has little to do with 
expectations and is nearly independent of future action 
of any agent involved.

\subsection{Potential prominence of promises}
Our belief in restricted strong non-obligationism  
implies that conceptually separating promises from obligations is 
a starting point of our work. Viewed as utterances promises must further be
separated from other utterances such as informative utterances, questions, predictions,
and commands. 

If an agent $A$ expresses  a fact, say $F$, meant to be informative for an audience, the similarity 
with $A$ promising $F$ is significant. However, when stating $F$, agent $A$ may not be able,
or have been able in the past, to
exercise its influence to bring about that $F$ holds true. 
While it is always acceptable that an agent acts in such a way
that its promises are kept, the same is not always valid for informative statements.

For instance: an agent telling an employer that ``he will not be present tomorrow because of a flu'' need
not be absent if the flu fails to hit as expected. Such a message implies no single action that increases its 
plausibility. What may get lost when viewing informative utterances as promises is objectivity. This provides
a clear separation, and at the same time it indicates that in systems that do not measure or assess agent 
objectivity it is unproblematic to treat each informative utterance as a promise.

When agent $A$ asks question $q$ to another agent, say $B$, in the scope of an audience, then the 
corresponding utterance need not be considered a promise. Indeed for understanding the notion
of a question one has no need of promises. In other words as concepts promises are not prior to questions.
Nevertheless it is quite plausible to replace the
question by an utterance where $A$ promises $B$ some reward upon $B$ 
providing an answer to $q$.

Finally, if agent $A$ issues a command $c$ to agent $B$ in scope $S$, then like with questions as just mentioned, 
in order to understand what is going one has no need of the concept of a promise. Having said that, however,
again it is plausible that $A$'s command is replaced by a promise that $A$ will provide some reward if $B$ acts as if it
puts command $c$ into effect. Suggesting this replacement is plausible in particular if one has no interest
in obligations that might result from the issuing of a command.

 We find that promises stand out as the utterances from which all other communications can
be derived. This gives promissory utterances some conceptual prominence of other utterances.

\subsection{For and against the primacy of obligations}
Invariably, however promises are linked with obligations. 
By the sheer weight of tradition,
obligations dominate discussions of behaviour. Our object is to position promises as a major
tool for the specification and the regulation of behavior. For that reason we make an attempt to 
compare promises and obligations as conceptual tools for distributed systems design.

\subsubsection{In favour of obligations}
We begin with listing some arguments in favor of obligations as regulatory mechanism
when compared with promises.
\begin{enumerate}
\item Some people might think that a promise is an obligation because it seems
to create one, and might therefore be considered equivalent to that obligation.
(This is a version of obligationism to which we oppose.)

\item Obligations are a well known concept from deontic logic. There is an advantage to to reducing
the less well-known concept of promises to one that has been studied
for more than fifty years. (This is true, but it implies no more than that promises are worth studying.)

\item Obligations have a formal status in state laws and regulations. There is no such
public body of promises. (Promises are a dynamic phenomenon concurrent with autonomous action, listing
promises globally and statically is not plausible.)

\item Many obligations give rise to promises which occur in the process of fulfilling
an obligation. E.g. the cat must get fed while owner is on holiday,
the owner is obliged to get the cat fed (by law forbidding cruelty to
animals). A friend promises to help in the fulfillment of the obligation.  (This is true but it does not
imply that obligations are prior to promises in general.)
\end{enumerate}

\subsubsection{Against obligations}
When balancing obligations and promises as a regulatory mechanism for distributed systems
significant arguments in favor of promises can be mentioned.
\begin{enumerate}
\item Obligations have their own complex logic, 
that is on philosophical grounds some combinations
of obligations are less plausible. 
Seemingly simple deontic  logic still captures (or hides) a
wealth of moral dilemmas and ethical mysteries. 
We refer to Holbo~\cite{Holbo2002}
for a strategies for dealing with inconsistent obligations.
\item Much more than obligations promises provide a tool for 
the self-organization of a community of
autonomous agents. At any stage, and concerning any future cooperative activity, 
agents can start exchanging promises thereby creating an increasingly 
complex and functional promise graph which may eventually provide a basis for usable 
expectation assessments. 
\item Once issued a promise admits a dynamically maintained equilibrium between the 
expectation of the promise being kept and  with credibility of and trust in the promiser. 
Such balancing tradeoffs are absent with obligations. 
\item It is easy to imagine a significant plurality of promises constituting a  promise graph related to a 
single planned activity. It is counterproductive to explain such promises in terms of equally 
many obligations.
 \item Suppose one has the concepts of promise available,
and now reflects upon obligations. An agent $P_{issue:ob}$ issuing an 
obligation with body $b_{ob}$ to all agents $Q$ in a scope $S$,
might be understood as simultaneously promising all members $Q$ of $S$ that $P_{issue:ob}$ will act in such
a way as if $P$ had received a promise with body $B_{ob}$ from $Q$.

This is a reasonable explanation of what may happen when issuing an obligation and it explains why
obligations seem to be complex entities or events from the perspective of promises.

\item If a future promise (e.g. the promise to feed the cat in the future) is
in fact a deception then this falsifies the necessity of a
relationship between promises and obligations.  In other words, all intentions occurring as 
apparent intentions in
promises cannot be induced obligations because some promises can be deceptions
and these cannot be (easily) understood as induced obligations.

\item Similarly, not all promises are about future actions, so there cannot be an implied
obligation capturing its essence for all promises. E.g. I promise that the cat got fed. 
Indeed the owner
might actually be displeased that the cat was fed if it was supposed to be dieting. 

\item There is a dissimilarity between promises and obligations. They are
quite different entities.  Obligations may cause promises and promises
may cause obligations, but promises have a physical reality as events in
space and time, whereas obligations do not.  Obligations are at a
different level of abstraction altogether.

\item Promises are made on a voluntary basis. For obligation however, the
concept of voluntarity is almost irrational. In any case it might be voluntary to
imply an obligation on someone else, but engaging in a promise you may face an
involuntary obligation or a voluntary one. Voluntarity is therefore natural for
promises but is quite problematic for obligations.

\item Promise announcement constitutes positive extensions of user behaviour, 
whereas obligations primarily
constitute a negative constraint on the degrees of freedom of the obliged party.

\end{enumerate}

\subsubsection{Promises precede obligations; locality precedes globality}
If one would choose between promises and obligations, it seems
abundantly clear that promises are the simpler concept, or perhaps the more
primitive concept. It is hard to imagine a world without promising while one 
may imagine a world without obligations. 

In addition the
concept of a promise seems more natural in the technological world:
since computers cannot feel ethical responsibility, the reduction to
promises to obligations seems to be neither philosophically
satisfactory nor technically correct.

So far we have argued that promises are different, simpler and can be
analyzed independently of obligations. There is one more point that is
of principal practical importance.  Promises are {\em local}
constructions, whereas obligations are {\em non-local}. The source of
a promise is localized in a single entity that has all of the
information and self-control to be available to resolve conflicts and
problems with multiple promises.  The sources of obligations however
are distributed amongst many individuals and the obliged party does
not have the access to resolve the conflicts without maintaining a
voluntary dialogue with all of these multiple parties.

From a practical viewpoint, obligations are simply less effective at
reducing uncertainty because they tend to increase uncertainty not
reduce it. Indeed, obligations can be inconsistent, but promises
cannot. More precisely: consistency of promises is a matter that can
be verified at the level of sources only. Promises made by different
agents cannot be inconsistent if promise bodies are dealing with actions or
states of affairs about which the promiser has an overriding control against
other agents. If different agents issue conflicting promises at least one of these
fails to have the expected degree of control. In any case once such 
conflicting promise are noticed the trust in both issuers needs reconsideration,
and a decrease in trust in at least one of the promise agents is likely to occur.

Preferably then one would not use obligation as a coordination
principle if a mechanism based on promises can be used
instead. 

\section{Conclusions}

We have restricted our attention to static scenarios in this work, as
the matter of change is a complex one. An account of how promises
arrive, persist and are removed again is forthcoming. Some work has
already been done in this area, 
however, in~\cite{papromise} and in \cite{sirimace2007}
but scope for embellishment is vast, as is the number applications for
the concept of promises. In the latter reference, the matter of {\em
organization} is related to promises, as a form of cooperation between
individuals or autonomous agents. 

Although we have avoided describing specific and detailed scenarios or
using the notion of agent autonomy in this work, one of us has written
extensively on the viewpoint that promises represent a basis for a sound
engineering discipline, somewhat analogous to an ``atomic theory'',
where arbitrary systems can be reduced to independent agents that issue
promises about their behaviour~\cite{burgessdsom2005}. This sets about
a process of documentation of independent properties that can be
helpful in detailing one's understanding of observed
phenomena. Indeed, in computing in particular there is a genuine case
to be made for viewing a computer as a number of independent
electrical components that make quite specific promises to one
another.

Given the role promises have in influencing certainty, an
exploration of the relationship to causation and fault propagation
networks seems fruitful. Similarly the benefits in economics are
obvious. 

Outside computing undoing promises from their obligation generating
effects can be a helpful to strengthen the role of promises as an
organizational principle. This may provide a fruitful direction for 
management science research.

Without attempting to suggest applications in any field, we have
argued for the usefulness of promises as an independent and practical
concept, whether in philosophy, economics or technology. We have
compared promises to the more usual idea of obligations and have
concluded unequivocally that promises are a simpler theoretical notion
and a more practical tool than obligations in the reduction of
an agent's uncertainty about the behaviour of other agents.

We show that intentions, promises, and commitments can be introduced and to
a significant extent explained in
the absence of an understanding of obligations. Furthermore, although
it seems to be a common view that obligations are a more primitive
concept than promises, our paper suggests the contrary.  Promises need
not be viewed merely as proxies for obligations; if promises give rise
to obligations, this can in fact be studied purely in an exposition
based on promises, intentions and commitments.  Indeed more often than
not promises are made by agents who would not be able to explain the
extent to which their promises might lead to obligations or not, or to
what extent such obligations would be more significant than the
promises from which they arose.


\end{document}